# Ultra-low-crosstalk Silicon Switches Driven Thermally and Electrically

Peng Bao[1], Chunhui Yao[1,2], Chenxi Tan[1], Alan Yilun Yuan[1], Minjia Chen[1], Seb J. Savory[1], Richard Penty[1], Qixiang Cheng[1,2*]

1. Electrical Engineering Division, Department of Engineering, University of Cambridge, UK
2. GlitterinTech Limited, Xuzhou, 221000, China

**Abstract**

Silicon photonic switches are widely considered as a cost-effective solution for addressing the ever-growing data traffic in datacenter networks, as they offer unique advantages such as low power consumption, low latency, small footprint and high bandwidth. Despite extensive research efforts, crosstalk in large-scale photonic circuits still poses a threat to the signal integrity. In this paper, we present two designs of silicon Mach-Zehnder Interferometer (MZI) switches achieving ultra-low-crosstalk, driven thermally and electrically. Each switch fabric is optimized at both the device and circuit level to suppress crosstalk and reduce system complexity. Notably, for the first time to the best of our knowledge, we harness the inherent self-heating effect in a carrier-injection-based MZI switch to create a pair of phase shifters that offer arbitrary phase differences. Such a pair of phase shifters induces matched insertion loss at each arm, thus minimizing crosstalk. Experimentally, an ultra-low crosstalk ratio below −40 dB is demonstrated for both thermo-optic (T-O) and electro-optic (E-O) switches. The T-O switch exhibits an on-chip loss of less than 5 dB with a switching time of 500 μs, whereas the E-O switch achieves an on-chip loss as low as 8.5 dB with a switching time of under 100 ns. In addition, data transmission of a 50 Gb/s on–off keying signal is demonstrated with high fidelity on the E-O switch, showing the great potential of the proposed switch designs.

**Introduction**

Recent years have seen the explosive growth of data traffic in both intra- and inter-datacenter networks, driven by the proliferation of data-intensive applications such as artificial intelligence (AI), cloud computing, and live streaming[1]. This escalating demand for optical interconnects has propelled advancements in optical switching technologies, taking advantage of their low power consumption, low latency, high bandwidth, and the elimination of optical-electrical conversions[2]. It is widely recognized that the future datacenter network architectures would be greatly benefited from optical circuit switching technologies, together with electrical packet switching[3]. To date, a variety of optical switches have been demonstrated based on different material systems and fabrication techniques, such as, MEMS[4,5], liquid crystal[6], lithium niobate (LiNbO$_3$)[7], III-V compounds[8,9], silicon nitride (SiN)[10,11], and silicon-on-insulator (SOI)[12,13]. Among these, the SOI platform is distinguished by its compact footprint, high-efficiency phase tuning, and most importantly, the compatibility of the CMOS process for mass-production.

Silicon switches typically rely on interferometric structures, such as micro-ring resonators (MRRs) or MZIs that are actuated by either T-O or E-O phase-shifting effect. In contrast to the MRR-based switches that are usually specialized for wavelength-selective routing, MZI switches offer broader bandwidths, relaxed fabrication requirements, and simplified control schemes[14]. Yet, their performance still falls short for practical adoptions, mainly on the aspects of insertion loss and coherent crosstalk. There are viable solutions on solving the switch loss by introducing additional gain blocks, via either monolithically or heterogeneously integrations, such as flip-chip bonding[15,16] and micro-transfer printing[17]. The crosstalk, on the other hand, represents a trickier issue to tackle, given the inevitable fabrication variations, inherent deficiency in switch topologies, and the most-challenging electro-absorption loss in E-O designs. This is particularly pronounced for large-scale switches, where the cumulative effect of crosstalk is more severe, considerably jeopardizing signal integrity[18].

Generally, the switch crosstalk is bounded by the fabrication imperfections in 3-dB couplers, where a slight deviation from the 50:50 splitting ratio could largely deteriorate the crosstalk. For example, 2% and 5% deviations could degrade the crosstalk ratio to below −28 dB and −20 dB, respectively[19]. For E-O switching, the situation gets worse due to the inherent electro-absorption loss on carrier injections. Current solutions for addressing these issues include employing a dilated topology to eliminate first-order crosstalk[20], implementing extra MZI stages as variable splitters[21,22], and applying nested MZI structures to improve power balance[23–25]. These approaches, however, all trade against the number of components, and thus resulting in larger device footprint and higher complexity. Thus, these demonstrations are more applicable to modest port count devices.

In this paper, we present two designs of silicon switches driven both thermally and electrically with ultra-low crosstalk. Customized strategies from the component level to the topology level are seamlessly combined with minimized system complexity. In specific, we demonstrate T-O and E-O switch elements with ultra-high extinction ratios, incorporating a dual-stage design to correct splitting-ratio errors and exploiting the self-heating effect in the E-O designs for loss balancing across MZI arms. These optimized switch elements are then applied to establish multi-port switch fabrics, being strategically deployed in the stages that are most vulnerable to crosstalk, in the partially dilated double-layer network (DLN) topology. Experimental results illustrate that both T-O and E-O switches in the scale of 8×8 achieve superior crosstalk ratios below −40 dB at the center wavelength of 1550 nm. This, to the best of our knowledge, sets a new record for the E-O switch fabrics. Additionally, the optical transmission of a 50 Gb on-off key (OOK) signal is performed on the E-O switch with high fidelity, demonstrating great potential for high-throughput optical interconnects.

## Results

### Design of T-O and E-O Switches

As schematically shown in Fig. 1(a), both the 8×8 T-O and E-O switches are assembled in a DLN topology that comprises 5 stages with a total of 64 switch elements. In such a topology, only the switch elements in the central stage encounter the first-order crosstalk, setting a limitation to the circuit[26]. Therefore, we implement dual-MZI switch elements in the central stage as ultra-low-crosstalk 2×2 cells, while the remaining four stages utilize regular MZIs as 1×2 or 2×1 cells, as shown by the insets in Fig. 1(a). This effectively suppresses the circuit crosstalk and reduces the total number of switch cells. Each single-MZI 1×2 and 2×1 cell comprises a pair of T-O or E-O phase shifters driven in a push-pull manner, while the dual-MZI 2×2 cell includes an additional MZI element with T-O phase shifters, serving as a variable power splitter. Thus, its coupling ratio can be complementarily matched with the that of the output combiner for mitigating the fabrication variation. It should be noted that the second MZI of the E-O 2×2 cell employs a doped phase shifter pair with distinct lengths and doping profiles, denoted as PS 1 and PS 2, respectively. This configuration aims to offer an arbitrary phase difference while maintaining an identical insertion loss on each MZI arm. Both 8×8 switches are optimized for TE mode operation at a wavelength of 1550 nm and utilize MMIs to enable a broadband operation. Figure 1(b) and 1(c) provide detailed schematics of the T-O and E-O phase shifters, respectively. The T-O phase shifter consists of a 300 μm TiN micro-heater, which is surrounded by deep air trenches for thermal isolation. As for the E-O phase shifter pair, the waveguide is laterally sandwiched between p- and n-doped regions. In specific, the E-O PS 1 features a length of 500 μm and contains heavily doped regions only, while the E-O PS 2 has a much shorter length of 100 μm and incorporates lightly doped regions in between the waveguide and the heavily doped regions. For both phase shifters, the p- and n-doped regions are separated by 2μm to lower the driving voltage and maintain isolation from waveguide mode. Additionally, they are equipped with a metal heater only for phase correction.

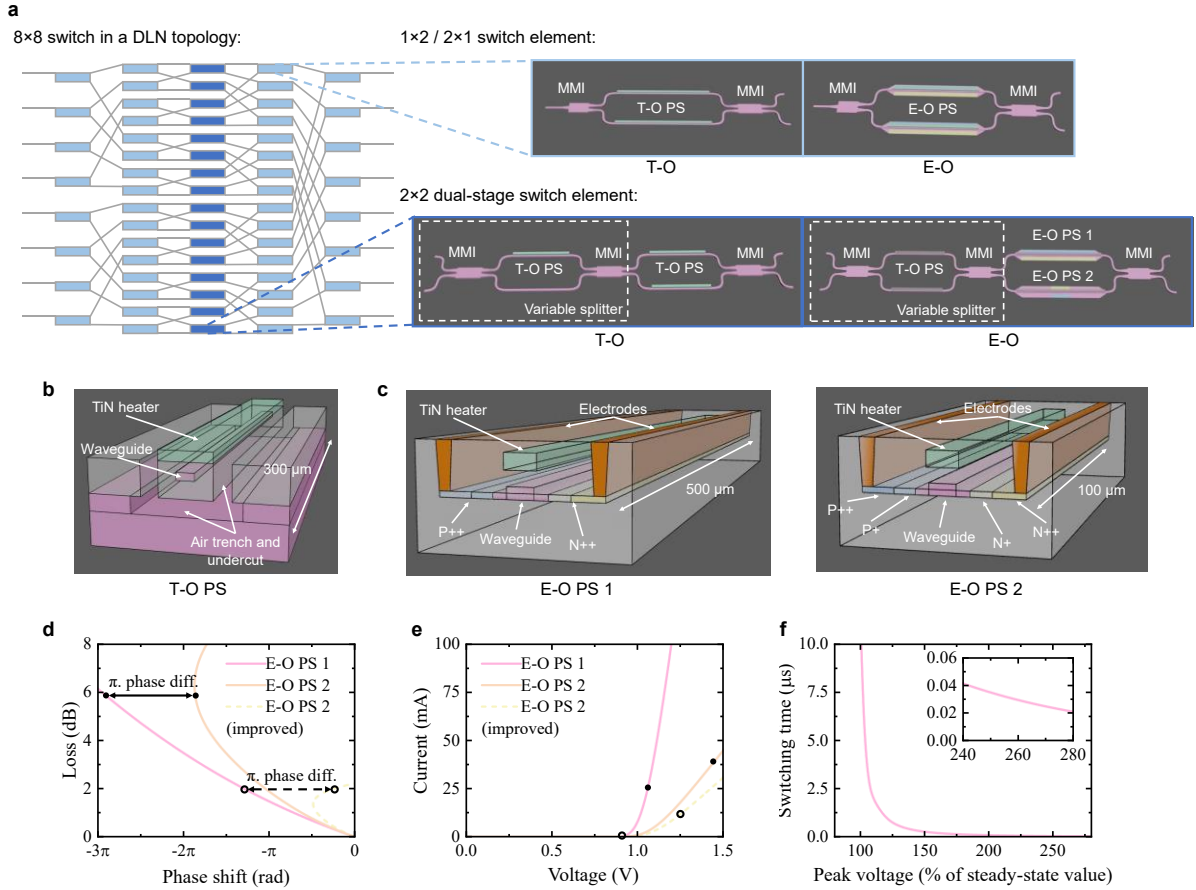

**Figure 1 | Design of the 8×8 T-O and E-O switches.** (a) The 8×8 DLN switch topology. Insets show the structure of the 1×2 and 2×2 MZI switch elements. (b) Schematic of the T-O phase shifter. (c) Schematic of the E-O differential phase shifter pair. (d) Insertion loss under different phase shift setting. Dots and circles on the curves represent the loss and phase for each phase shifter when the switch element is configured in the bar state. (e) I-V characteristics of the E-O phase shifters, with dots and circles indicating the voltage and current for each phase shifter in the bar state. (f) Switching time of the E-O differential phase shifter pair, varying with pulse peak voltages.

In typical E-O switches, the phase modulation is realized by injecting free carriers from the doped regions into the waveguides, where the concentration increase of free carriers lowers the waveguide's refractive index via free carrier dispersion (FCD). This, however, introduces unwanted loss due to the free-carrier absorption (FCA) that creates power imbalances between the MZI arms, leading to incomplete interference and, consequently, crosstalk. Moreover, Soref's equations[27], which tie the change in refractive index to the absorption coefficient for silicon, indicate that the induced phase change is always accompanied by a fixed amount of loss, leaving little room to individually manipulate phase change and FCA loss. To break this bound, we enhance the self-heating effect, which offsets the phase change caused by the FCD effect but retains the FCA loss. This heat generates from both Joule heating from carrier currents and by carrier recombination, whose impact intensifies with driving current. Consequently, we develop a pair of doped phase shifters with different lengths. The shorter phase shifter (E-O PS 2), which aims to match the loss of the longer shifter as per Soref's equations, is designed to experience more severe self-heating effect. The inclusion of lightly doped regions further exacerbates this impact, as carrier recombination rates are higher in these regions. It undergoes both FCD and self-heating effects, offers nearly counterbalanced phase change but FCA-induced loss. Meanwhile, the longer phase shifter (E-O PS 1) leverages the FCD effect only, providing fast phase shifting with FCA-induced loss. Such a phase shifter pair can thus operate differentially to achieve a balanced loss and an arbitrary overall phase difference to trigger switching with minimal crosstalk, as illustrated in Fig. 2(d). The detailed I-V characteristics for each phase shifter are shown in Fig. 2(e). The current cell design is simulated to have an overall insertion loss of 6 dB, with a power consumption of approximately 100 mW. It is expected that the insertion loss and power consumption can be further reduced to 2 dB and around 20 mW, respectively, by further shortening the length of PS 2 and optimizing the position of its electrodes to enhance self-heating[28].

Notably, in the proposed E-O 2×2 cell, the self-heating effect inevitably impairs the switching speed. Simulations indicate that both the rise and fall times triggered by the self-heating effect are approximately 12 μs. In this work, pulse excitation technique[29] and differential control schemes[30] are employed to enhance the switching speed. Specifically, an additional excitation pulse is applied on top of the step-on signal to overdrive it and accelerate the temperature rise. Conversely, when cooling it, a pulse is applied to the phase shifter on the other MZI arm to rapidly heat it, thereby reducing the temperature differential between the two arms and the resultant phase difference. Figure 2(f) confirms that by applying excitation pulses with sufficiently high voltages, the switching time of the proposed device can be reduced to less than 100 nanoseconds.

**Experimental characterization**

Figures 2(a) and 2(b) showcase the microscope images of the T-O and E-O switch chips, respectively, with each occupying a footprint of $6.5 \times 5.7$ mm$^2$ and $7.8 \times 6.1$ mm$^2$. Figures 2(c) and 2(d) illustrate these chips after electrical and optical fanout. The T-O and E-O chip respectively have 162 and 276 electrical pads in total and both are wire-boned to customized PCBs for electrical

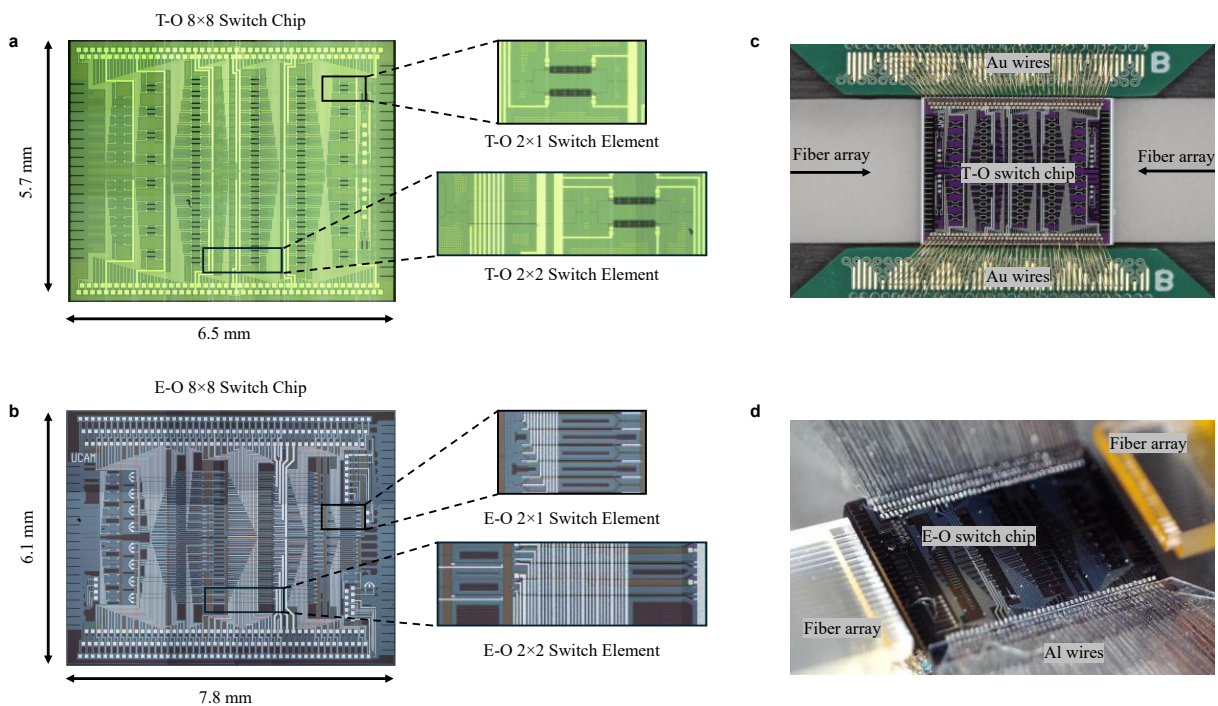

**Figure 2 | The microscope image and package of the switches.** (a) The microscope image of the T-O switch. (b) The packaged T-O switch. (c) The microscope image of the E-O switch. (d) The package E-O switch.

connections.

To characterize the switch performance, we first perform switch calibrations. Figure 3(a) illustrates the transmission of a T-O 1×2 cell in accordance with the driving power, where approximately 2 mW of power is required to achieve a π phase shift. The 2×2 cell, on the other hand, consumes 1 mW to toggle between cross and bar states, as shown in Fig. 3(b). For the E-O switch, its 1×2 switch element necessitates about 1 mW for a π/2 phase shift, as per Fig. 3(c), while this number goes up to around 100 mW for the 2×2 cell since the self-heating effect is enhanced. Figure 3(d) presents the 2×2 switch element's transmission response to bias current variations. The tilting pattern rising from the bottom left to the top right indicates the increasing self-heating effects in E-O PS 2. This agrees well with the simulations.

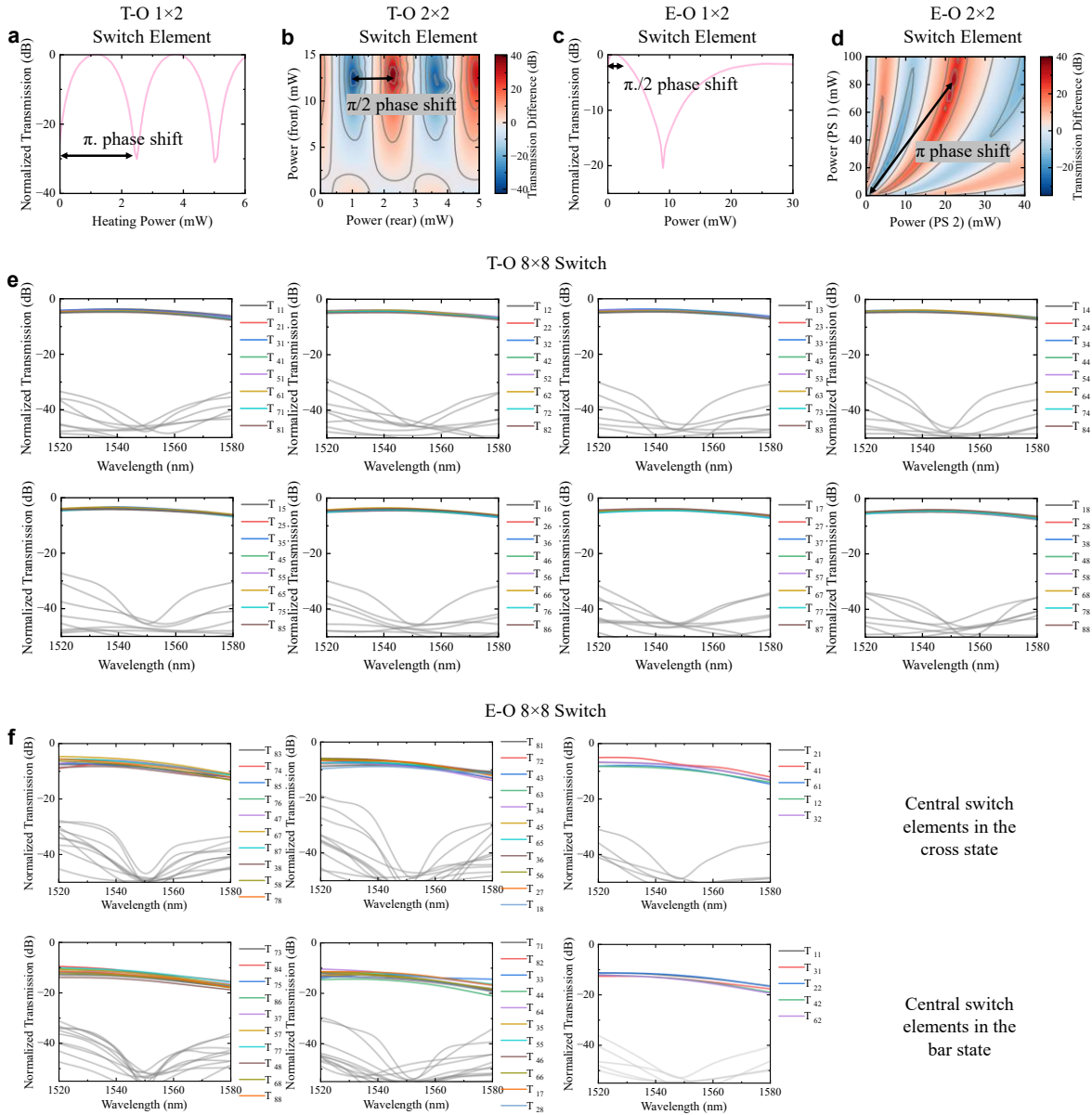

**Figure 3 | The performance of the T-O and E-O switches**. (a-d) The transmission of the 1×2 and 2×2 T-O and E-O switch element against the driving power, respectively. (e-f) The transmission spectra of all routing paths for the 8×8 T-O and E-O switches, respectively. The colored lines represent the transmission from specific input ports to their corresponding output ports, with grey lines indicating crosstalk leaked to undesired outputs.

Figure 3(e) shows the spectral responses of different optical paths for the T-O switch over the wavelength window between 1520 nm and 1580 nm. The on-chip propagation loss ranges from 4 to 5 dB, with slight variations due to differences in path length and number of waveguide crossings. The crosstalk is below −40 dB at the central wavelength and stays under −30 dB across a bandwidth greater than 40 nm. Figure 3(f) presents the transmission spectra of the E-O switch, with certain optical paths unavailable also due to faulty E-O phase shifters. As can be seen, the on-chip propagation loss measures around 8.5 dB, when the middle-stage 2×2 switch elements are set to the cross state, and changes to below 14 dB when they are switched to the bar state. Under both circumstances, our E-O switch maintains a crosstalk ratio below −40 dB at the wavelength of 1550 nm, which keeps below -30 dB across a bandwidth of about 10 nm.

To characterize the transient response of the switches, we utilize square wave signals of either 0.1 kHz or 1 kHz from an arbitrary waveform generator (AWG), which are amplified by an electrical amplifier and applied to drive the T-O or E-O switch. The output signal from the switch circuit is amplified using an erbium-doped fiber amplifier (EDFA), which is filtered for detection. The rise and fall times are then measured using an oscilloscope. Figure 4(a) and 4(b) presents the recorded rising and falling temporal waveforms of the 1×2 and 2×2 T-O switch elements, respectively, both featuring a switching time of approximately 500 μs. Similarly, Fig. 4(c) to 4(f) illustrates the temporal waveforms of the 1×2 E-O switch element with a regular E-O phase shifter, showing a rise time of 10 ns and a fall time of 60 ns. As for the 2×2 E-O switch element with differential phase shifters, using a conventional driving method—where square wave signals are applied simultaneously to both shifters—results in a rise and fall times of 2 μs and 10 μs, respectively, as shown in Fig. 4(d). Such a reconfiguration speed results from the time required for the waveguide to reach thermal equilibrium, owing to the inducement of self-heating. A pulse excitation technique is subsequently adopted combining with the differential control to drive the two phase shifters. In specific, when heating up, we accelerate the process by applying an additional high-energy pulse, while at cooling down, we rapidly heat the other arm to quickly reduce the temperature difference, effectively mimicking fast cooling. Figures 4(e) details the applied driving signals, where PS 2 receives a 5 V pulse with a duration of 40 ns at the rising edge, and PS 1 receives a 4 V, 50 ns pulse at the falling edge. Here both amplitude and duration of these pulses are carefully optimized to shorten the switching time while preventing any over-drive damage. Figure 4(f) depicts the measured waveforms, showing a rise time of 40 ns and a fall time of 100 ns, respectively. Note that the switching speed can be further improved by applying higher-energy pulses, but a thorough investigation on the burn-down threshold at different doping levels has to be conducted. Thus, we opt to employ a moderate pulse energy as a balanced choice, since a switching speed of <100 ns is widely considered sufficient for datacenter applications[31].

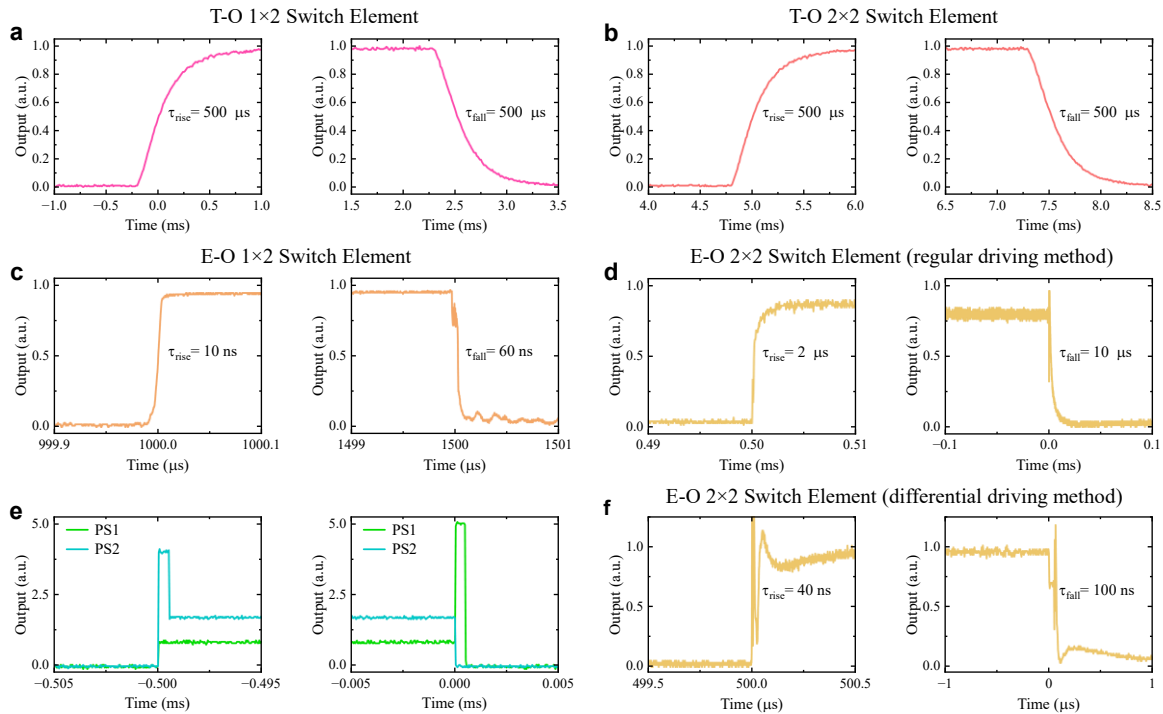

**Figure 4 | Transient responses of the T-O and E-O switch elements.** (a-b) Measured rise and fall time of the 1×2 and 2×2 T-O switch elements, respectively. (c) Measured rise and fall time of the 1×2 E-O switch element. (d) Measured rise and fall time of the 2×2 E-O switch element under regular driving. (e) The driving signals of the differential control scheme. (f) Measured rise and fall time of the 2×2 E-O switch element after applying differential control scheme.

**Data transmission performance**

To verify the switching capabilities of our designs, a data transmission experiment is conducted using the 8×8 E-O switch as an example. Figure 5(a) outlines the measurement setup, where the transmitted signals are generated by modulating a continuous wave laser at a wavelength of 1559 nm with a high-speed electro-absorption modulator. This modulator is driven by a 50 Gb/s OOK signal produced by an arbitrary waveform generator (AWG). The modulated optical signal is adjusted to TE polarization and coupled into the switch chip. After routing, the output power is attenuated by a variable optical attenuator (VOA) before sampled by an oscilloscope. Offline digital signal processing (DSP) techniques are employed to calculate the bit-error-rate (BER). Figure 5(b) illustrates the analytical fit of the BER results for both the back-to-back (BtB) transmission and the transmission involving the switch. Meanwhile, Fig. 5(c) presents the eye diagrams for the eight routing paths at a received optical power of -5 dBm. The results reveal that, at a fixed BER of $10^{-2}$, the inclusion of the E-O switch in the system introduces a power penalty of less than 0.8 dB due to post-amplification, demonstrating high fidelity.

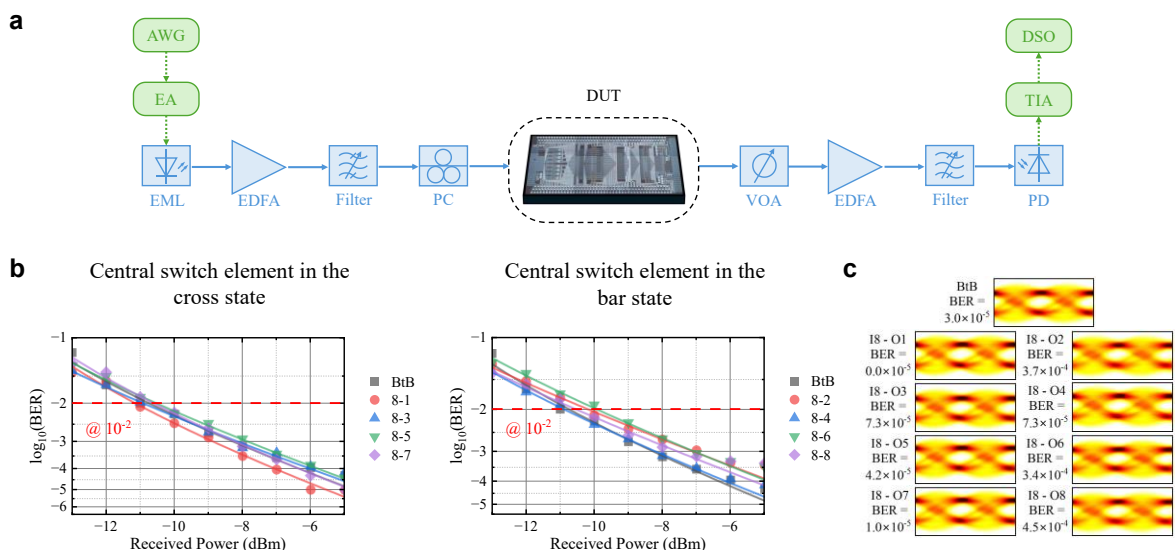

**Figure 5 | High-speed data transmission experiment for the E-O switch.** (a) Schematic of the BER measurement setup; (b) BER results of the back-to-back (B2B) transmission and the eight routing paths from input I8; (c) Eye diagrams of the B2B transmission and the eight routing paths from input I8.

## Discussion and conclusion

To highlight the demonstrated switches, this work is placed in context with other demonstrations of low-crosstalk switch fabrics on the SOI platform, as shown in Table 1. By employing a partially-dilated DLN topology with a dual-MZI design in the central stage, our T-O switch achieves superior crosstalk suppression, with values below −40 dB at the central wavelength and under −30 dB across a 40 nm range, matching the performance of fully-dilated strictly non-blocking topologies. Thanks to excellent thermal isolation, each T-O switch element consumes only about 2 mW to toggle between the cross and bar states. Note that while the thermal isolation impacts switching speed, this can still be mitigated using pulse excitation techniques, leaving the switching time constrained only by the heat propagation from the heater to the waveguide.[29] On the other hand, the proposed E-O switch also realizes a crosstalk level below −40 dB, being >10 dB better than state-of-the-art results. This advantage is attributed to the usage of differential phase shifter pair, which addresses the FCA-induced power imbalances through a straightforward structure.

Moreover, our switch design features the scalability towards larger port counts. The inherent characteristics of the DLN topology ensure that low crosstalk can be maintained by applying the proposed differential phase shifter design in the central stage, regardless of switch size. To manage the increased control complexity in large-scale switches, a column-row addressing method can be implemented to selectively actuate the necessary phase shifters[32], significantly reducing the number of electrical control signals.

In conclusion, we present two generations of ultra-low-crosstalk silicon switches driven thermally and electrically. Utilizing a DLN topology with customized switch elements, both switches achieve crosstalk levels below −40 dB, setting a new performance benchmark in the field. Moreover, the T-O switch delivers an insertion loss of <5 dB with a 500 μs switching time, while the E-O switch offers an insertion loss as low as 8.5 dB and a switching speed of <100 ns. The

**Table 1. Performance comparison with state-of-the-art low-crosstalk switch fabrics**

| Category | Work | Port count | Topology | Blocking Characteristic | IL [dB] | CT [dB] | Speed μs/ns | Footprint [mm²] |
|---|---|---|---|---|---|---|---|---|
| T-O | [33] | 8×8 | S&S | SNB | ~4.0 | <-30 (min), <-16 (over 80 nm) | 250 | 8.0×8.0 |
| | [34] | 8×8 | PA-S&S | SNB | 2.5-4.0 | <-40 (min), <-30 (over 90 nm) | N/A | 8.5×4.5 |
| | [35] | 16×16 | Beneš | RNB | 4.2-6.2 | <-30 (over 10 nm) | 22 | 7.0×3.6 |
| | [36] | 32×32 | Dilated-Beneš | RNB | ~15-~40 | ~35 | 70/1400 | 12.0×12.0 |
| | This work | 8×8 | DLN | SNB | 4.0-5.0 | <-40 (min), <-30 (over 40 nm) | 500 | 6.5×5.7 |
| E-O | [37] | 8×8 | DLN | SNB | 5.7-8.7 | <-30 (over 7 nm) | 10 | 12.0×7.0 |
| | [20] | 16×16 | Customized dilated topology | SNB | 12-19 | <-30 (over 40 nm) | 6 | 12.6×12.6 |
| | This work | 8×8 | DLN | SNB | 8.5-14 | <-40 (min), <-30 (over 15 nm) | 100 | 7.8×6.1 |

IL: Insertion loss; CT: Crosstalk; RNB: Rearrangeably non-blocking; SNB: Strictly non-blocking

optical transmission of a 50 Gb OOK signal is demonstrated, showing a high fidelity. Overall, the proposed switch fabrics can bring about a breakthrough in the scalability of E-O switch fabrics, and thus may find wide applications in the next-generation datacenter networks.

## Methods

### Simulation of the differential E-O phase shifter pair

ANSYS Lumerical CHARGE and MODE software tools are used to conduct comprehensive electrical, thermal and optical simulations of the differential E-O phase shifter pair. Initially, a coupled heat and charge transport solver is set up to model the heat propagation and carrier distribution within the phase shifter under various bias voltages. This solver self-consistently solves the drift-diffusion equations with Poisson's equation and the heat transport equation using the finite-element method (FEM). The profiles generated from this process are subsequently imported into a finite-difference eigenmode (FDE) solver, which calculates the resultant changes in the effective refractive index $\Delta n_{eff}$ and FCA loss. Finally, $\Delta n_{eff}$ is translated into phase shift $\Delta \phi$ using the following equation:

$$\Delta \phi = \frac{2\pi}{\lambda} \Delta n_{eff} \Gamma L, \tag{1}$$

where $\lambda$ is the wavelength of the signal, $\Gamma$ is the confinement factor, and $L$ is the length of the phase shifter.

To estimate the switching time of the E-O phase shifter, we perform a transient analysis. The process begins by determining the steady-state current density and recombination rate of the phase shifter with the MZI in the bar state, using a charge transport solver. These parameters are then used to calculate the heat generation within the device by the following equations:

$$Q = Q_n + Q_p + Q_R, \tag{2}$$

$$Q_{n,p} = \mathbf{J}_{n,p} \cdot \mathbf{E}_{n,p}, \tag{3}$$

$$Q_R = q(E_g + 3kT)R, \tag{4}$$

where $\mathbf{J}_{n,p}$ is the current density, $\mathbf{E}_{n,p}$ is the electric field, $q$ is the electron charge, $E_g$ is the bandgap energy, $k$ is the Boltzmann constant, $T$ is the temperature, and $R$ is the net recombination rate. During the simulation of the switching process, this generated heat is treated as a source within the heat transport solver, which operates in transient mode. The activation and deactivation of this heat source are controlled by a step signal incorporating a 10 ns rise time that follows a logarithmic profile, mimicking the exponential change in carrier concentration during actuation. This setup allows for the extraction of temperature variations over time in the waveguide core, from which the switching time is determined. Throughout these simulations, the ambient temperature is maintained at 300°K around the simulation region.

### Chip fabrication and packaging

The T-O and E-O switches chips are fabricated via an Advanced Micro Foundry (AMF) silicon photonics multi-project wafer (MPW) run, employing 193-nm deep ultraviolet (DUV) lithography capable of achieving feature sizes down to 140 nm. The process is based on 220-nm silicon-on-insulator (SOI) technology, with a 3-µm buried oxide (BOX) layer and a 2-µm top cladding layer.

Two 20-channel single mode fibre arrays (SMFAs) are used to couple to and from the two facets of each chip, with a coupling loss of around 4 dB per facet. A thermistor is attached to the bottom of the chip, and a thermoelectric cooler (TEC) is placed underneath to form a negative feedback mechanism, facilitating the temperature stabilization during testing.

An automatic electrical control plane has been developed to drive the switches. Control signals generated by a computer are first sent to a microcontroller unit (MCU), then transmitted to high-resolution digital-to-analog converters (DACs) via a serial peripheral interface (SPI). The voltages from the DACs are subsequently amplified and loaded onto the chip. Specifically, current-output DACs are used for the differential E-O phase shifters to precisely control the current flow, while voltage-output DACs are employed for the T-O phase shifters and regular E-O phase shifters.

**Chip calibration**

To calibrate the switches, we employ a C-band tunable laser set at an output power of 0 dBm as the light source, while the output signals from the switches are collected by an 8-channel optical power meter. In the calibration of each switch element within the T-O switch, we sweep the bias voltage applied to its phase shifter and monitor the optical power at the related output ports. We then determine the voltages for setting these switch elements in the cross and bar states by identifying the bias voltages that maximize or minimize the output power difference. Additionally, the central dual-MZI switch elements necessitate the fine-tuning of both phase shifters. To avoid the time-consuming two-dimensional sweep, we employ a gradient descent algorithm to efficiently search for the optimal bias voltages.

The calibration process for the E-O switch mirrors that of the T-O switch. We calibrate each $1\times2$ / $2\times1$ switch element at the bilateral stages by sweeping the bias voltages on the heaters to identify the 0 and $\pi$ phase biases. After setting the heater's voltage to the quadrature point to achieve a $\pi/2$ phase bias for the push-pull configuration, we adjust the voltages on the E-O phase shifters. This creates a look-up table (LUT) for maximizing or minimizing transmission power, forming a control map for both heaters and E-O phase shifters that defines the cross and bar states of the $1\times2$ / $2\times1$ MZI cells. For the central $2\times2$ switch elements, we initially use a gradient descent algorithm to identify the optimal bias voltages for their heaters to pre-bias these elements to the cross state. Once all switch elements are set to the cross state, we refine the voltage combinations for the differential E-O phase shifters to set their bar state. Consequently, the control LUT for the entire E-O switch is obtained.

**Data availability**

For the purpose of open access, the author has applied a Creative Commons Attribution (CC BY) license to any Author Accepted Manuscript version arising. Data underlying the results presented in this paper are available at https://doi.org/XXXX/CAM.XXXX.

**Acknowledgement**

This work was supported by the UK EPSRC Programme Grant QUDOS (EP/T028475/1), TRANSNET (EP/R035342/1), Communication Hub TITAN (EP/X04047X/1 and EP/Y037243/1), UK EPSRC CEPS CDT (EP/S022139/1), the European Union's Horizon Europe Research and Innovation Program under Agreement 101070560 (PUNCH), and the European Union's Horizon 2020 research and innovation program, project INSPIRE (101017088). C.Y. and C.T. thank the scholarship provided by Cambridge Trust and China Scholarship Council. The authors thank Dr. Giuseppe Talli and Dr. Maxim Kuschnerov for the support in the project.

## Author contributions

P.B. conceived the switch design, performed the optical simulations, and drawn the chip layout. P.B. performed the characterization of switch performance and analyzed the data with C.Y. and M.C.'s help. P.B., C.T. and A.Y.Y. designed and conducted the data transmission with M.C.'s assistance. P.B. and C.Y. drafted the manuscript, with Q.C., C.T., S.J.S. and M.C.'s input. R.P. and Q.C. supervised the project.

## Conflict of interests

The authors declare no conflict of interests.